\documentstyle[12pt,epsfig]{article}

\renewcommand{\title}[1]{{\Large\bf\mbox{}\\\medskip#1\bigskip\medskip\\}}
\renewcommand{\author}[1]{{\large #1\smallskip\\}}
\newcommand{\address}[1]{{\em #1\medskip\\}}

\def\be{\begin{equation}}
\def\ee{\end{equation}}
\def\bea{\begin{eqnarray}}
\def\eea{\end{eqnarray}}
\def\ba{\begin{array}}
\def\ea{\end{array}}
\def\a{\alpha}
\def\b{\beta}
\def\c{\gamma}

\def\G{\Gamma}
\def\0{$\Gamma_0$}
\def\o{\omega}
\def\p{\phi}

\def\t{\theta}

\def\s{\sigma}

\def\l{\lambda}

\begin{document}
\begin{center}

\title{Density of the Fisher zeroes for the Ising model}


\author{Wentao T. Lu and F. Y. Wu}
\address{Department of Physics\\
 Northeastern University, Boston,
Massachusetts 02115}

\begin{abstract}
  The density of the Fisher zeroes, or zeroes of the partition function
in the complex temperature plane, 
is determined for the Ising model in zero field
as well as in a pure imaginary field $i\pi/2$.
Results are given for
 the simple-quartic, triangular, honeycomb, and the
kagom\'e lattices.
   It is found that the density  diverges logarithmically at  points along its loci. 
\end{abstract}

\end{center}



\section{Introduction}
In the analyses of lattice models in statistical mechanics
such as the Ising model, the partition
function is often expressed in  the form of a polynomial in variables such as
the external magnetic field and/or the temperature.  Since properties of a polynomial
are completely determined by its roots,
a knowledge of the zeroes of the partition function 
yields all thermodynamic properties of the system.  Particularly,
if the  zeroes lie on a certain locus, a knowledge of its 
density distribution along the locus is equivalent to the obtaining of
the exact solution of the problem.

For the Ising model with ferromagnetic interactions,  we have the remarkable
Yang-Lee circle theorem \cite{yanglee} which 
states that all  partition function zeroes
lie on the unit circle $|z|=1$
in the complex $z=e^{2L}$ plane, where $L$ is the reduced external magnetic field
$(we set kT=1)$.  However, the density of the Yang-Lee zeroes on the unit circle, a knowledge
of which is equivalent to  solving the Ising model in a nonzero magnetic
field, is known
only for the Ising model in one dimension.

Fisher \cite{fisher} has proposed
that it is also meaningful to consider  partition function zeroes
in the complex temperature plane.
  Indeed, he showed that for the zero-field Ising model on the simple
quartic lattice with nearest-neighbor reduced interactions $K$, 
the
partition function zeroes lie on  two circles
\be
|\tanh K \pm 1 | = \sqrt {2} \label{twocircles}
\ee
in the thermodynamic limit.
He further showed that the known logarithmic singularity of the 
specific heat follows from  the fact that
the density  vanishes linearly near the real axis. 
 Subsequently, the Fisher loci
has    been determined for the infinite triangular lattice  \cite{steph},
and for {\it finite}  simple-quartic 
lattices which are self-dual \cite{luwu}.  
Stephenson \cite{steph1} has also evaluated the density 
distribution on the circles
in terms of a Jacobian. However, the explicit expressions of the density
function of the Fisher zeroes do not appear to have been 
heretofor evaluated.

In this paper we complete the picture by evaluating the density function.
We deduce the explicit expressions for the density of Fisher zeroes 
for the  simple-quartic, triangular, honeycomb, and
kagom\'e lattices.  Density of the Fisher zeroes 
for the Ising model
in a pure imaginary field $L=i\pi/2$ are also obtained.

\section{The simple-quartic lattice}
It is well-known that
the bulk solution of spin models with short-range interactions
is independent of the boundary conditions.  For the Ising model 
on the simple-quartic lattice, we shall take
a particular boundary condition  
introduced by Brascamp and Kunz \cite{bk} for which the location of 
the Fisher zeroes is known for any {\it finite} lattice.  This permits us to take a
a well-defined and unique bulk limit, thus avoiding a difficulty encountered
in the consideration of the Ising model on a torus \cite{steph1}.

Consider an $M\times 2N$ simple-quartic lattice with cylindrical boundary conditions
in the $N$ direction and fixed boundary conditions along the two edges
of the cylinder.  The  $2N$ boundary spins on each of the two edges of the cylinder
have fixed fields $\cdots ++++++ \cdots$ and $\cdots +-+-+- \cdots$, respectively. 
This is the Brascamp-Kunz boundary condition \cite{bk}.
Brascamp and Kunz showed that the partition function of this Ising model
is precisely
\be
Z_{M,2N}(K)=2^{2MN} \prod_{1\leq i \leq N} \prod_{1\leq j \leq M} \biggl[
1+z^2-z (\cos \t_i + \cos \p_j)\biggr], \label{bkpart}
\ee
where 
\be
z=\sinh 2K, \hskip 1cm
 \t_i=(2i-1)\pi/{2N}, \hskip 1cm \p_j = j\pi/(M+1).\label{finitezero}
\ee
 The per-site ``free energy" in the bulk  limit is then evaluated as
 \bea
f &=& \lim_{M,N\to \infty} {1\over {2MN}} \ln \ Z_{M,2N}(K) \nonumber \\
&=& {1\over 2} \ln (4z) +{1\over 8\pi^2}\int_{-\pi}^{\pi}d\t\int_{-\pi}^{\pi}d\p\ 
\ln[z+z^{-1}-(\cos\t+\cos\p)] \nonumber \\
&=& {1\over 2} \ln (4z) +{1\over 2\pi^2}\int_{0}^{\pi}d\t\int_{0}^{\pi}d\p\ 
\ln[z+z^{-1}-2\cos u\cos v],  \label{freef}
\eea    
where $u=(\t+\p)/2, v=(\t-\p)/2$ and we have made use of the fact that the  integrands
are $2\pi$-periodic.

The partition function (\ref{bkpart}) has  zeroes   at
the $2MN$ solutions of
\be
z+z^{-1} =  \cos \t_i + \cos \p_j,\ \ \ 1\leq i \leq N; 1\leq j \leq M.
\label{unitc}
\ee
The following lemma and corollaries are now used to determine the loci of the zeroes:

\medskip
\noindent
{\it Lemma}: The regime   $-2\leq z+z^{-1} \leq 2$
of the complex $z$ plane, where $z+z^{-1} =$ real,
is the unit circle $|z|=1$.

 \medskip
\noindent
Proof:
The Lemma follows from the fact that, by writing
$z=re^{i\t}$, we have 
\be
z+z^{-1}=\biggl(r+{1\over r}\biggr) \cos \t +i\biggl(r-{1\over r}\biggr)\sin \t,
\ee
so that $z+z^{-1}=$ real implies either $r=1$ or $\t =$ integer $\times \pi$.
In the latter case we have $|z+z^{-1}|= |r+r^{-1}|>2$, which contradicts
the assumption, unless $r=1$.  It follows that we have always $r=1$, or $|z|=1$.
Q.E.D.

\medskip
\noindent
{\it Corollary 1}: The regime 
  $-a\leq z+z^{-1} \leq b$, where $a,b>2$,
$z+z^{-1} =$ real, of the
complex $z$ plane  is the union of the unit circle $|z|=1$ and  segments
$z_-(-a)\leq x \leq z_+(-a)$ and $z_-(b)\leq x \leq z_+(b)$ of the real
axis, where $z_\pm (b) = (b\pm\sqrt{b^2-4})/2$. 

\medskip
 \noindent
{\it Corollary 2}: The regime 
 $-a\leq z+z^{-1} \leq b$, where $a,b>2$, $z+z^{-1} =$ real, 
 of the
complex $z$ plane, is the regime $|w|=1$ in the complex $w$ plane, where
$w$ is the solution of the equation
\be
w+w^{-1} ={4\over {a+b}} \biggl(z+z^{-1} + {{a-b}\over 2}\biggr).
\ee

Corollary 1 is established along the same line as in the proof of the lemma,
and Corollary 2 is a consequence of the lemma since, by construction, we have 
$-2\leq w+w^{-1} \leq 2$.  

\medskip
Returning to the partition function (\ref{bkpart}),
since the right-hand side of (\ref{unitc}) is real and lies in $[-2, 2]$,
 it follows from the Lemma
 that the $2MN$ zeroes of 
(\ref{bkpart}) all lie on the unit circle $|\sinh 2K|=1$, a result 
which can also be obtained 
 by simply setting the argument of the logarithm in the 
bulk free energy  (\ref{freef}) equal to zero.
The usefulness of this simplified procedure
  has   been pointed out by Stephenson and Couzens \cite{steph}
for the Ising model on a torus.  But since the zeroes are not easily determined
in that case when the lattice is finite, they termed the argument  as ``hand-waving".
   Here, the argument is made rigorous by the use of the Brascamp-Kunz boundary
condition.  From here on, therefore, 
  We shall 
adopt this simpler approach
 in all  subsequent considerations.

\medskip
We now proceed to determine the density of the  zero distribution.
Let the number of zeroes in the interval $[\a,
\a+d\a]$ be  $2MN g(\a) d\a$ such that 
\be
\int_{0}^{2\pi}g(\a)d\a=1 
\ee
and
\be
f= {1\over 2}\ln (4z)+ \int_{0}^{2\pi}d\a \ g(\a)\ \ln(z-e^{i\alpha}).\label{limit}
\ee
It is more convenient to consider the function $R(\a)=\int
_0^\a g(x)dx$
where  $2MN R(\a)$ gives the total number of zeroes in the
interval $[0,\a]$ such that
 \be
g(\a) ={d\over{d\a}} R(\a).  \label{totalroot}
\ee

\medskip
On the circle $|z|=1$ writing $z=e^\a$ and setting the argument of the
logarithm in  the third line of (\ref{freef}) equal to zero, 
we find   $\a$  determined by
\be
 \cos \a = \cos u \cos v, \hskip 1cm 0 \leq \{u,v\}\leq \pi. \label{xy1}
\ee
Now if $\a_i$ is a solution, so are $-\a_i$ and $\pi-\a_i$, hence we have
the symmetry
 \be
g(\a)=g(-\a) = g(\pi -\a). \label{gsymmtry}
\ee
It is therefore sufficient to consider only $0\leq \{\a,u,v\} \leq \pi/2$.

\medskip

\begin{figure}[htbp]
\center{\rule{5cm}{0.mm}}
\rule{5cm}{0.mm}
\vskip -.8cm
\epsfig{figure=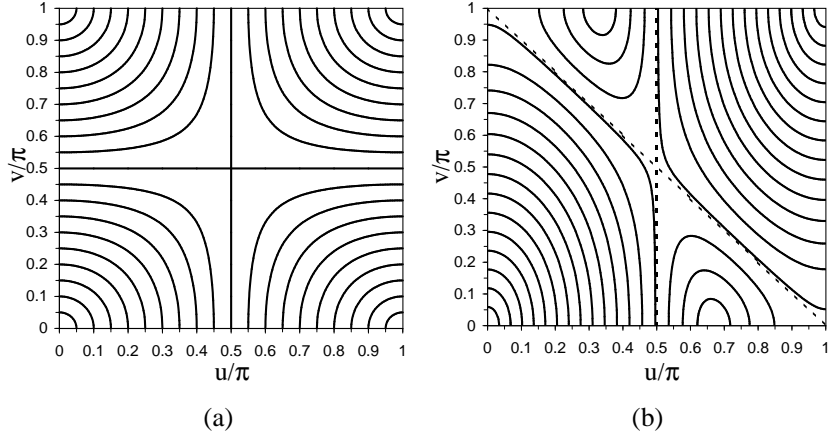,height=3.5in,angle=0}
\vskip -3.2cm
\caption{Constant-$\a$ contours in the $u$-$v$ plane.
(a). The contour (\ref{xy1}) for the simple-quartic lattice. 
Straight lines correspond to
$\a=\pi/2$. (b). The contour (\ref{xytri}) for the triangular lattice.
 Broken
lines correspond to $\a=2\cos^{-1}(1/3)$.}
\label{fig:fig1}
\end{figure}

The constant-$\a $ contours of \ref{xy1} are constructed in Fig. 1(a) and are seen
to be  symmetric about the lines $u,v= \pm\pi/2$ in
each of the 4 quadrants. 
Now  from (\ref{finitezero})  we see that zeroes are distributed uniformly in the 
$\{\t,\p\}$-, and hence the $\{u,v\}$-plane.
It follows that  $R(\a)$ is precisely the area of the region
\be
 \cos \a > \cos u \cos v, \hskip 1cm 0 \leq \{\a,u,v\} \leq \pi/2, \label{xy}
\ee
normalized to $R(\pi/2)=1/4$.  
This leads to the expression  
\be
R(\a)={\frac{1}{\pi^2}}\int_0^{\a}\cos^{-1}\Bigg({\frac{\cos\a}{
\cos x}}\Bigg)dx. \label{densitya}
\ee
Using (\ref{totalroot}) and
  after some reduction, we obtain the following explicit expression for the 
density of zeroes, 
\be
 g(\a)=R'(\a)=
  {\frac{|\sin\a|}{\pi^2}}K(\sin\a),\label{density-sq}
\end{equation}
where $K(k)=\int_0^{\pi/2}dt(1-k^2\sin^2t)^{-1/2}$ 
 is the complete elliptic integral of the first kind.
 The density (\ref{density-sq}), which possesses an unexpected
logarithmic divergence at    $\a=\pm\pi/2$, is plotted in Fig. 2(a).
 For small $\a$, we have $g(\a) \approx |\a|/2\pi$.
As pointed out by Fisher \cite{fisher}, it is this linear behavior
at small $\a$  which leads to the logarithmic 
divergence of the specific heat.
 
\medskip
We can also deduce the density of zeroes on the two Fisher circles
(\ref{twocircles}) which we write as
\be
\tanh K \pm 1 = \sqrt 2 e^{i\t}.  \label{2cir}
\ee
The angles $\a$ and $\t$ are related by,
\be
e^{i\a} =\pm \Big( {{\sqrt 2 \mp e^{-i\t}}\over {\sqrt 2 \mp e^{i\t}}}\Big) \label{at}
\ee
so that the mapping from $\a$ to $\t$ is 1 to 2.
This leads to the result
\be
g(\t) = {{g(\a)}\over 2} \Bigg| {{d\a}\over {d\t}} \Bigg|.
\ee
    Let the density of zeroes be $g_\pm(\t)$
 for the two circles (\ref{2cir}).  Then, using (\ref{at}) we find
\bea
g_+(\t) &=& g_-(\pi- \t) \nonumber \\
&=&   
\biggl({k\over \pi^2}\biggr)\Bigg|
{1-\sqrt{2}\cos\t\over 3-2\sqrt{2}\cos\t}\Bigg|K(k), \label{densityb}
\eea
where
\be
k={2|\sin\t|(\sqrt{2}-\cos\t)\over 3-2\sqrt{2}\cos\t}.
\ee
The density (\ref{densityb}) is plotted as Fig. 2(b).
Note that the divergence in the density distribution in (\ref{densitya})
is removed in (\ref{densityb}).  The points  $\a=\pm \pi/2$ in
$g(\a)$ is mapped onto the points $\t=\pm(\pi/2 \pm \pi/4)$ in $g_\pm (\t)$.
We have $g_+(\pi/4)=g_-(3\pi/4)=0$, and for small $\t$ we find
\be
g_\pm (\t) = \Big({{3\pm 2\sqrt 2}\over {\pi}} \Big)|\t|.
\ee
Here, again, the linear behavior of
$g_+(\t)$ at $\t=0$  leads to the logarithmic singularity of the specific heat.
 
\medskip
It is also of interest to consider zeroes of the Ising model 
 in the Potts variable $x=(e^{2K}-1)/\sqrt 2$.
In the complex $x$ plane it is known \cite{chenhuwu} that
the partition function zeroes are on two unit circles centered at
$x=1$ and $x=-\sqrt 2$. 
We find the density along the two circles  to
be, respectively, $g_-(\t)$ and $g_+(\t)$.
 
\begin{figure}[htbp]
\center{\rule{5cm}{0.mm}}
\rule{5cm}{0.mm}
\vskip -0.9cm
\epsfig{figure=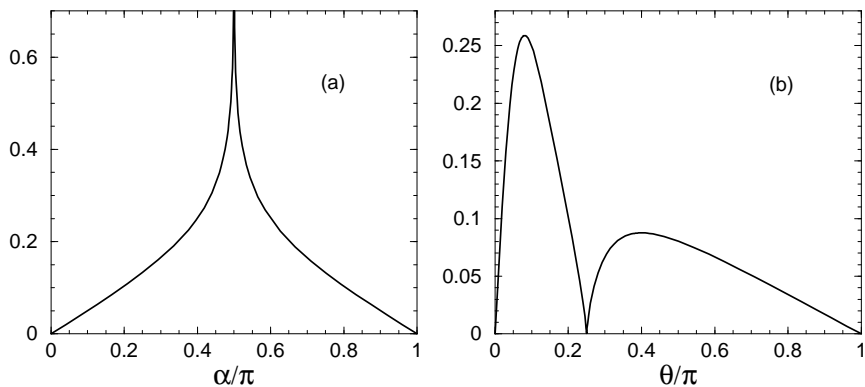,height=4.5in,angle=-90}
\vskip -.2cm
\caption{Density of  partition function zeroes 
for  the simple-quartic lattice.
(a).  $g(\a)$ given by (\ref{density-sq}). 
 (b).  $g_{+}(\t)$ given by (\ref{densityb}).}
\label{fig:fig2}
\end{figure}

\section{The triangular lattice}
For the triangular Ising model with nearest-neighbor interactions $K$, the
free energy assumes the form \cite{hou,wannier}
\bea
 f&=&C+{1\over 8\pi^2}\int_{-\pi}^{\pi}d\t\int_{-\pi}^{\pi}d\phi
\ln\biggl[z+z^{-1}+1
-[\cos\t+\cos\p+\cos(\t+\p)]\biggr] ,\nonumber \\
&=&C +{1\over {2\pi^2}}
\int_{0}^{\pi}du\int_{0}^{\pi}dv
\ln \Big[z+z^{-1} +2 -2\cos u(\cos u+\cos v)\Big],
  \label{trif}
\eea
where $C=[\ln (4z)]/2$, $z=(e^{4K}-1)/2$, 
and we have  introduced variables $u=(\t+\p)/2,v=(\t-\p)/2$.
Now the value of the sum of the three cosines
in (\ref{trif}) lies between $-3/2$  and $3$.  It then follows from Corollary
1 that in the complex $z$ plane the zeroes lie on the union of
the unit circle $|z|=1$ and the line segment $[-2, -1/2]$ of the real
axis, a result first obtained by Stephenson and Couzens \cite{steph}.

The density of the zero distribution can now be computed in the same manner
as described in the preceding section.   
  For $z$ on the unit circle we write $z=e^{i\a}$.
Then  $\a$ is determined by
\be
\cos \a = -1+\cos u(\cos u+\cos v), 
\hskip 1cm 0 \leq \{u,v\}\leq \pi, \label{xytri}
\ee
and $R(\a)$ is the area of the region
\be
\cos \a > -1+\cos u(\cos u+\cos v). 
\ee
Clearly, we have the symmetry $g_{\rm cir}(\a)=g_{\rm cir}(\pi-\a)$ and we need
only to consider $0\leq \a \leq \pi$.

From a consideration of  the constant-$\a$ contours of \ref{xytri}
shown in Fig. 1(b), we 
obtain after some algebra  the result
\be
g_{\rm cir}(\a) 
={|\sin\a|\over {\pi^2\sqrt{A(\a)}}}K(k), \label{densityc}
\ee
where $A(\a) = (5+4\cos\a)^{1/2}$ and
 \bea
k^2&=&  F[A(\a)] \nonumber \\
F(x) &\equiv& {1\over 16}\Big({3\over x}-1\Big)(1+x)^3.
\eea
 Particularly, for small $\a$,
 we find $g_{\rm cir}(\a)\approx |\a|/2\sqrt{3}\pi$.

In a similar fashion we find, on the line segment $z\in [-2,-1/2]$,
we write $z=-e^{\l}$ and obtain
\be
g_{\rm line}(\l)={|\sinh \l|\over \pi^2k\sqrt{B(\l)}}K(k^{-1}),
\hskip 1cm  -\ln 2\leq \l\leq \ln 2 ,\label{densityl}
\ee
where $B(\l) = [5-4\cosh \l]^{1/2}$ and
 \be
k^2=  F[B(\l)] .
 \ee
In contrast to the case of the simple-quartic lattice, the density of zeroes
is everywhere finite. Specifically, we have
  $g_{\rm cir}(\pi)=g_{\rm line}(0)=0$,
   and $g_{\rm line}(\pm \ln2)=\sqrt{3}/2\pi$.
 The densities (\ref{densityc}) and (\ref{densityl})  are plotted in Fig. 3.

\begin{figure}[htbp]
\center{\rule{5cm}{0.mm}}
\rule{5cm}{0.mm}
\vskip -0.9cm
\epsfig{figure=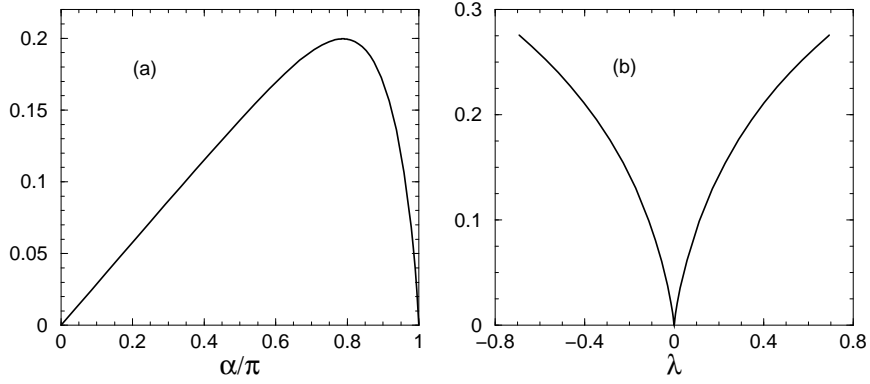,height=4.5in,angle=-90}
\vskip -.2cm
\caption{Density of partition function zeroes 
for the triangular lattice.
(a). $g_{\rm cir}(\a)$ given by (\ref{densityc}).
  (b). $g_{\rm line}(z)$ given by (\ref{densityl}).} 
 \label{fig:fig3}
\end{figure}

\medskip
Matveev and Shrock \cite{ms} have discussed zeroes of the triangular
Ising model in the complex $u=e^{-4K}$ plane, for which
 the zeroes are distributed on the union of the circle
\be
u={1\over 3}(2e^{i\a}-1), \qquad -\pi<\a\leq\pi,
\ee
and the line segment
\be
-\infty <u \leq -{1\over 3}.
\ee
Using our results we find the  respective densities
\be
g_{\rm cir}(\a)={|\sin\p|\over {9\pi^2}}[C(\a)]^{7/2}K(k),
\ee
where $C(\a) =3(5-4\cos\p)^{-1/2}$, $k^2= F[C(\a)]$,
 and 
 \be
g_{\rm line}(u)=\Bigg|{(1+u)(1-3u)\over 4\pi^2u^2(1-u)^2k\sqrt {D(u)}}\Bigg|\ 
K(k^{-1}),
\ee
where $D(u)= \sqrt {(1+3u)/u(1-u)}$ and  $k^2=F[D(u)]$.
At the end point we have $g_{\rm line}(-1/3)=9\sqrt{3}/8\pi$.

\medskip
The density of zeroes assumes a simpler form if we use
 Corollary 2 to map all zeroes onto a unit circle in 
 the complex $w$ plane, where  $w$ is root of the quadratic equation
  \be
w+w^{-1} ={4\over 7} \biggl(z+z^{-1} + {3\over 4}\biggr),
\ee
and $z=(e^{4K}-1)/2$.
For $w$ on the unit circle, we write $w=e^{i\a}$ and
 find in analogous to (\ref{xy}) that $R(\a)$ is the area of the region
 \be
\cos\a>{1\over 9}\Big[8\cos u(\cos u+\cos v)-7\Big].
\ee
Using the contours  shown in Fig. 1(b), we obtain
\begin{eqnarray}
R(\alpha ) &=&{\frac{1}{\pi ^{2}}}\int_{0}^{\phi _{0}}\cos^{-1} \Bigg[{\frac{
9\cos \alpha +7}{8\cos {\phi }}}-\cos {\phi }\Bigg]d\phi ,
\hskip 1cm \alpha \in
[0,\alpha _{0}]  \nonumber \\
&=&{\frac{1}{2}}-{\frac{1}{\pi ^{2}}}\int_{\phi _{0}}^{\phi _{1}}\cos^{-1} 
\Bigg[{\frac{9\cos \alpha +7}{8\cos {\phi }}}-\cos {\phi }\Bigg]d\phi ,\quad
\alpha \in [\alpha _{0},\pi ],
\end{eqnarray}
where $\a_0=2\cos^{-1}(1/3)$ and 
\begin{eqnarray}
\phi _{0} &=&\cos^{-1} \Bigg[{\frac{3}{2}}\cos {\frac{\alpha }{2}}-{\frac{1}{2}}
\Bigg]   \nonumber \\
\phi _{1} &=&\pi -\cos^{-1} \Bigg[{\frac{3}{2}}\cos {\frac{\alpha }{2}}+{\frac{1
}{2}}\Bigg],\quad {\rm for}\quad \cos {\frac{\alpha }{2}}\le {\frac{1}{3}}.
\end{eqnarray}
Note that we have $R(\alpha _{0})=3/8$, $R(\pi )=1/2$.
 
Finally, using (\ref{totalroot}), we obtain 
\begin{eqnarray}
g(\alpha)&=&{\frac{9\sin\alpha}{8\pi^2}}\int_0^{\phi_0} {\frac{d\phi}{\sqrt{
(\cos^2\phi-\cos^2\phi_0)(\Delta^2-\cos^2\phi)}}}, \hskip 0.8cm  \a\in [0,\a_0]  
\nonumber \\
&=&{\frac{9\sin\alpha}{8\pi^2}}\int_{\phi_0}^{\phi_1} {\frac{d\phi}{\sqrt{
(\cos^2\phi-\cos^2\phi_0)(\cos^2\phi_1-\cos^2\phi)}}},\quad \alpha\in
[\alpha_0,\pi], \label {gg}
\end{eqnarray}
where $\Delta=[1+3\cos({\alpha}/{2})]/2$.
After some manipulation and making use of 
 integral identities (\ref{I1}) and (\ref{I2}) derived in the Appendix,
we obtain 
\bea
g(\alpha)&=&{\frac{3\sqrt{3}}{8\pi^{2}}}\Big|\sin \alpha \Big| 
\sqrt{\sec {\frac{\alpha }{2}}}K(k),
\quad \alpha \in [0,\alpha _{0}]\nonumber \\
&=&{\frac{3\sqrt{3}}{8\pi^{2}}\Bigg|{\sin \alpha \over k }\Bigg| 
\sqrt{\sec {\frac{\alpha }{2}}    } }
K(k^{-1}),\quad \alpha \in [\alpha_{0},\pi ]  \label{density-tri}
\eea
where 
\begin{equation}
k^2=
{\frac{1}{16}}(\sec {\frac{\alpha }{2}}-1)\Big(1+3\cos {\frac{\alpha }{2}}
\Big)^{3}.
\label{tri-k}
\end{equation}
Note that  $g(\alpha)$ diverges logarithmically at $\alpha=\pm\alpha_0$.

\section{Simple-quartic Ising model in a field $i\pi/2$}
The two-dimensional Ising model can  be solved when there is an external
magnetic field $i\pi/2$. The solution for the simple-quartic lattice 
was first given by Lee and Yang \cite{leeyang} 
 and a rigorous derivation of which was given later by McCoy and Wu \cite{mw}.
In 1988 Lin and Wu \cite{lw} gave a
 general prescription for writing down the  solution of the Ising model
in a field
$i\pi/2$ by transcribing  the solution in a zero
field.
The most general known solution of the Ising model in a field $i\pi/2$
 is a model with a generalized checkerboard
type interactions \cite{wu86}.
   We consider in this section the case of the simple-quartic lattice. 

 For the simple-quartic lattice Lee and Yang \cite{leeyang} gave the
free energy in a field $i\pi/2$ as
\be
 f=i{\pi\over 2}+C
+{1\over 16\pi^2}\int_{-\pi}^{\pi}d\t\int_{-\pi}^{\pi}d\phi
\ln[z+z^{-1}+2-4\cos\t\cos\p], \label{pffield}
\ee
where $C= (\ln \sinh 2K)/2, z=e^{-4K}$.
Setting the argument of the logarithm in (\ref{pffield}) equal to zero
we have $-6\leq z+z^{-1}\leq 2$ and hence from
 Corollary 2 we see that in the complex $z$ plane zeroes of the partition
function lie on the unit circle $|z|=1$ and the line segment 
$-3-2\sqrt{2}\leq z  \leq -3+2\sqrt{2}$ of the real axis.

On the unit circle $|z|=1$ we write $z=e^{i\a}$ and find the density
\be
g_{\rm cir}(\a)={|\sin\a|\over 2\pi^2}K(k),\label{density5}
\ee
where
\be
k^2=(3+\cos\a)(1-\cos\a)/4.
\ee
On the line segment, we write $z=-e^{\l}$ with
$-2\ln(1+\sqrt{2})\leq \l \leq2\ln(1+\sqrt{2})$,
we find the density
 \be
g_{\rm line}(\l)={|\sinh \l|\over 2\pi^2}K(k),
\label{density6}
\ee
where
\be
k^2=(3-\cosh \l)(1+\cosh \l)/4.
\ee
At the end points we have $g_{\rm line}(\pm 2\ln(1+\sqrt{2}))=
1/\sqrt{2}\pi$.
The density functions (\ref{density5}) and (\ref{density6}) are plotted in Fig. 4.

\begin{figure}[htbp]
\center{\rule{5cm}{0.mm}}
\rule{5cm}{0.mm}
\vskip -0.9cm
\epsfig{figure=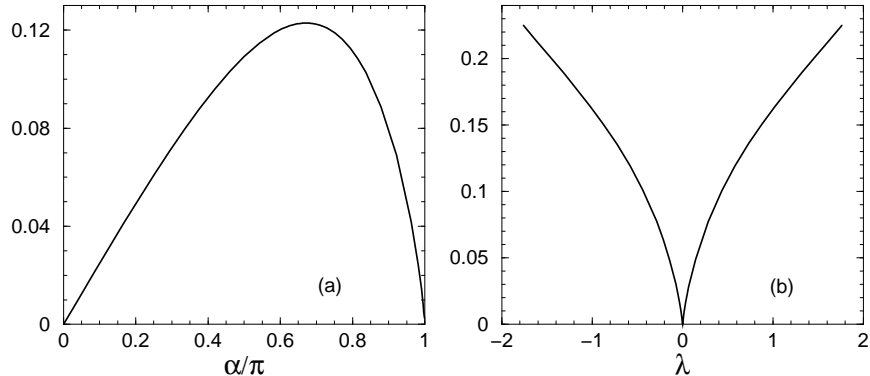,height=4.5in,angle=-90}
\vskip -.2cm
\caption{Density of partition function zeroes for
the simple-quartic lattice Ising model in a  field $i\pi/2$.
(a).  $g_{\rm cir}(\a)$ given by (\ref{density5}).
(b). $g_{\rm line}(z)$ given by (\ref{density6}).}
\label{fig:fig4}
\end{figure}

\section{Triangular Ising model in a field $i\pi/2$}
 The solution for the triangular model in a field $i\pi/2$ was first
obtained in \cite{lw} 
by applying a transformation in conjunction with  the solution of
a staggered 8-vertex model.   Here,
for completeness, we present an alternate and more direct
derivation.
 
\begin{figure}[htbp]
\center{\rule{5cm}{0.mm}}
\rule{5cm}{0.mm}
\vskip -0.8cm
\epsfig{figure=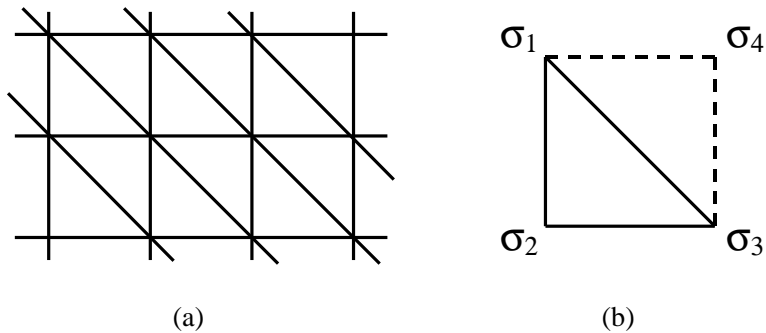,height=6.5in,angle=0}
\vskip -11.2cm
\caption{(a). The triangular lattice.
(b). A unit cell.}
\label{fig:fig5}
\end{figure}

Consider a triangular Ising lattice of $N$ sites whose sites are
arranged as shown in Fig. 5(a).
After making use of the identity
 $e^{i\pi\s/2}=i\s$, the partition function assumes the form
\be
Z_N
= i^N\sum_{\s_i=\pm 1} \prod_{\rm nn} e^{K\s_i\s_j} \prod_ j \s_j \label{trifield}
\ee
where the first product is over all nearest neighbors,
and the second product over all sites.
 Now  it is known that the  triangular Ising model can be mapped 
into an 8-vertex model on the dual of the square lattice \cite{fw}, also of $N$ sites.  
However, in order to properly treat the  
factor $\prod_j\s_j$ in  (\ref{trifield}), we need  to divide the 
$N$ ``cells" of the lattice, where a cell is shown in Fig. 5(b),
 into two sublattices, $A$ and $B$, and associate  two  $\s_j$'s
to each cell belonging to one
sublattice, say, $B$.  This permits us to rewrite (\ref{trifield}) as
\be
Z_N
= i^N\sum_{\s_i=\pm 1}\  \prod_{\rm cells}  W_{\rm stg}(\s_1, \s_2, \s_3, \s_4)
\ee
where $W_{\rm stg}(\s_1, \s_2, \s_3, \s_4)$ is a staggered Boltzmann  weight given by 
\bea
W_{\rm stg}(\s_1, \s_2, \s_3, \s_4) &=&  e^{K(\s_1\s_2+\s_2\s_3+\s_3\s_1)}
  \hskip 1.5cm {\rm for\>\>}A \nonumber \\
&=& (\s_1\s_2)e^{K(\s_1\s_2+\s_2\s_3+\s_3\s_1)} \hskip 0.5cm {\rm for\>\>}B.
\label{8vweight}
\eea
The 8-vertex weights are
 \bea
\{\o_1, \cdots, \o_8\}&=&\{e^{3K}, e^{-K},e^{-K},e^{-K},e^{-K},e^{-K},e^{-K},e^{3K}
\} \nonumber \\
\{\o_1', \cdots, \o_8'\}&=&\{e^{3K}, -e^{-K},-e^{-K},e^{-K},e^{-K},-e^{-K},e^{-K},-e^{3K}
\}.  \label{w}
\eea
Furthermore, from the mapping convention of  Fig. 1 of \cite{hlw}, we see that
the mapping between the spin and 8-vertex configurations is 2 to 1.
This leads to 
\be
Z_N=2 i^NZ_N(\{\o\}, \{\o'\}),
\ee
which is
an exact equivalence between  $Z_N$ and the 
partition function $Z_N(\{\o\}, \{\o'\})$ of the
staggered 8-vertex 
model.
 
Now the weights (\ref{w}) satisfy the free-fermion condition \cite{fw}
for which  $Z_N(\{\o\}, \{\o'\})$
has already been evaluated \cite{hlw}.
Using Eq. (19) of \cite{hlw} and after some reduction, one obtains the following
 expression for the per-site free energy,
\be
f=i{\pi\over 2}+C+{1\over 4\pi^2}\int_{0}^{\pi}d\t\int_{0}^{\pi}d\phi
\ln\Big[(1+e^{4K})^2 +4\cos\p(\cos\t+\cos\p)
\Big],
\ee
where $C=[\ln( 2\sinh 2K)]/2$.
As a result, the partition function zeroes are located at
\be
(1+e^{4K})^2=-4\cos\p(\cos\t+\cos\p), \hskip 1cm 0\leq\{\t,\p\}\leq \pi. \label{ddd}
\ee
It is therefore convenient to consider the $z=e^{4K}$ plane.
Since
\be
-8 \leq (1+e^{4K})^2\leq 1,
\ee
using the Lemma we find that
the zeroes are on the union of the segment
$-2\leq z\leq 0$ of the real axis
and  the line segment $z=-1+iy$, $-2\sqrt{2}\leq y\leq 2\sqrt{2}$.
The density of zeroes can be similarly determined.
On the  segment $z\in[-2,0]$ of the real axis, we find
\be
g(z)=\Bigg|{(1+z)\over 2\pi^2k\sqrt{E(z)}}\Bigg|\ K(k^{-1}),\label{tri-re}
\ee
where $E(z)=\sqrt{-z(2+z)}$ and $k^2=F[E(z)]$.  Particularly, 
we have $g(0)=g(-2)=1/\sqrt{3}\pi$ and $g_(-1)=0$.
On the line segment $z=-1+iy$, we find
\be
g(y)=\Bigg|{y\over 2\pi^2\sqrt{H(y)}}\Bigg|\ K(k),\label{tri-im}
\ee
where $H(y)=\sqrt{1+y^2}$ and $k^2 = F[H(y)]$.  Particularly, 
 we have $g(0)=0$ and
$g(\pm 2\sqrt{2})=1/\sqrt{6}\pi$.
These results are plotted in Fig. 6.

\begin{figure}[htbp]
\center{\rule{5cm}{0.mm}}
\rule{5cm}{0.mm}
\vskip -0.9cm
\epsfig{figure=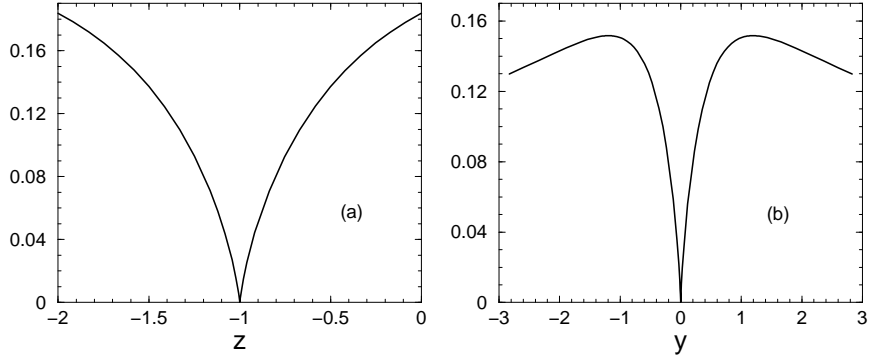,height=4.5in,angle=-90}
\vskip -.2cm
\caption{Density of partition function zeroes for
the triangular Ising model in a  field $i\pi/2$.
(a).  $g(z)$ given by (\ref{tri-re}).
(b). $g(y)$ given by (\ref{tri-im}).}
\label{fig:fig6}
\end{figure}

we remark that in
 the complex $x=e^{-4K}$ plane considered in \cite{ms}, the segment $-2\leq z\leq 0$
of the real axis maps onto $-\infty \leq x\leq -1/2$ while
the line segment $z=-1+iy$, $-2\sqrt{2}\leq y\leq 2\sqrt{2}$,
 is mapped onto  the circular arc  
$x^{-1}={1\over 2}(-1+e^{i\t})$, $\t_0=\tan^{-1}(4\sqrt{2}/7)\leq |\t|\leq \pi$.
The density of zeroes on the arc is found to be
\be
g_{\rm arc}(\t)={(1+\cos\t)^2\over 2\pi^2\sqrt{I(\t)}|\sin^3\t|}\ K(k),
\ee
where $I(\t)=[2(1+\cos\t)]^{1/2}/|\sin\t|$ and $k^2=F[B(\t)]$.
The densities at the end points of the arc are
$g_{\rm arc}(\pm\t_0)=3\sqrt{3}/2\sqrt{2}\pi$.

\section{The honeycomb and kagom\'e lattices}
The partition function of an Ising model on a planar lattice 
with interactions $K$ is proportional to the partition function on the dual
lattice with interactions $K^*$ \cite{wuwang}, where $K$ and $K^*$ are
related by
\be
e^{-2K^*} = \tanh K.
\ee
Consequently, their partition function zeroes coincide when expressed in terms
of appropriate variables.   
Now the honeycomb and triangular lattices are
mutually dual, it follows that for  the honeycomb
lattice  with interactions $K$, in the complex
 \be
z={1\over 2}(e^{4K^*}-1)=(\cosh 2K-1)^{-1}
\ee
plane, zeroes of the partition function coincides with those of the
triangular lattice partition function (\ref{trif}).

For the honeycomb  Ising model in an
external  field $i\pi/2$, the free energy can be obtained
from that in a zero field via a simple transformation \cite{lw,ms}.
Writing the partition function in the form of
(\ref{trifield}) and replacing the product $\prod_i\sigma_i$  by
$\prod_i\sigma_i^3$, it is clear that, besides the factor $i^N$, the
partition function is the same as that in a zero field with the replacement
\be
e^{K(\sigma_i\sigma_j-1)} \to (\sigma_i\sigma_j)e^{K(\sigma_i\sigma_j-1)} ,
\ee
or, equivalently, $e^{2K} \to -e^{2K}$.
It follows that in the complex
\be
z=(-\cosh 2K-1)^{-1}
\ee
plane, the zeroes coincide with those of the triangular lattice  partition
function (\ref{trif}).

The Ising model on the kagom\'e  lattice with interactions $K$
 can be mapped to  that
on an honeycomb lattice with interactions $J$,
by applying a star-triangle transformation
followed by  a spin decimation.
 The procedure, which is standard \cite{syozi} and
will not be repeated  here, leads to the relation
\be
e^{2J}=({e^{4K}+1})/ 2. 
\ee
As a result, we conclude that, in the
complex
\be
z=(\cosh 2J-1)^{-1}=2(1-\tanh 2K)/ \tanh^2 2K
\ee
plane,  zeroes of the kagom\'e partition function 
 coincides with those of the triangular lattice partition function
(\ref{trif}).
The evaluation of the   kagom\'e partition function 
in an external field $i\pi/2$ remains unresolved, however.

\section*{Acknowledgment}

Work has been supported in part by NSF grants DMR-9614170 and DMR-9980440.


\setcounter{section}{0}
\renewcommand{\thesection}{APPENDIX}
\begin{center}
\section{Two integration identities}
\end{center}
\setcounter{equation}{0}
\renewcommand{\theequation}{\Alph{section}\arabic{equation}}

In this Appendix we derive the integration
identities
\bea
I_1&=&\int_{0}^{\pi/2}{\frac{dt}{\sqrt{(1-a^{2}\sin^{2}t)(b^{2}+a^{2}\sin^{2}t)}}}
={\frac{1}{\sqrt{a^{2}+b^{2}}}}
K\Bigg(a\sqrt{{\frac{1+b^{2}}{a^{2}+b^{2}}}}\Bigg)
\label{I1}  \\
I_2&=&\int_{a}^{b}{\frac{dx}{\sqrt{(1-x^{2})(x^{2}-a^{2})(b^{2}-x^{2})}}}
={\frac{1}{b\sqrt{1-a^{2}}}}
K\Bigg({\frac{1}{b}}\sqrt{{\frac{b^{2}-a^{2}}{1-a^{2}}}}\Bigg), \label{I2}
\eea
 which do not appear to have previously been given.

To obtain (\ref{I1}), we expand the integrand using
 the binomial expansion 
\begin{equation}
(1-x)^{-\a}=\sum_{k=0}^{\infty}{\frac{(\a)_k}{k!}}x^k,
\end{equation}
where $(\a)_k=\a(\a+1)\cdots(\a+k-1)=\G(\a+k)/\G(\a)$, and 
carry out the integration term by term using the formula
\begin{equation}
{2\over \pi}\int_0^{\pi/2}\sin^{2m}t dt={\frac{({\frac{1}{2}})_m}{m!}} .
\end{equation}
This yields 
\begin{eqnarray}
I_1 &=&{\frac{\pi}{2b}}\sum_{j,k=0}^{\infty} {\frac{({\frac{1}{2}})_j
({\frac{1}{2}})_k({\frac{1}{2}})_{j+k} }{j!k!(j+k)!}}a^{2j}
\Big(-{\frac{a^2}{b^2}}\Big)^k 
\nonumber \\
&\equiv &{\frac{\pi}{2b}}F_1\Bigg({\frac{1}{2}};{\frac{1}{2}},{\frac{1}{2}};1;
a^2, -{\frac{a^2}{b^2}}\Bigg),
\end{eqnarray}
where the hypergeometric function of two variables is (cf. 9.180.1 of
\cite{gr})
\begin{equation}
F_1(\a;\b,\b';\c; x,y)=\sum_{j=0}^{\infty}\sum_{k=0}^{\infty} {\frac{
(\a)_{j+k}(\b)_j(\b')_k }{j!k!(\c)_{j+k}}}x^{j}y^k.
\end{equation}
This leads to the integration formula ({\ref{I1}) after making use of
the identity (cf. 9.182.1 of \cite{gr})
\begin{equation}
F_1(\alpha;\beta,\beta';\beta+\beta';x,y)=(1-y)^{-\a}F\Bigg(
\a,\b;\b+\b';{\frac{x-y}{1-y}}\Bigg),
\end{equation}
where $F$ is the hypergeometric function (cf. 9.100 of \cite{gr})
\be
F(\a,\b;\c; z)=\sum_{j=0}^{\infty} {\frac{
(\a)_{j}(\b)_j}{j!(\c)_{j}}}\ z^{j}\ ,
\ee
and the identity  (cf. 8.113.1 of \cite{gr})
\be
K(k)={\pi\over 2}F\Bigg({1\over 2},{1\over 2};1;k^2\Bigg).
\ee

The integral (\ref{I2}) is obtained by  introducing the change of   variable
 $x^2=(b^2-a^2)\sin^2t+a^2$, which yields
 \begin{equation}
I_2={\frac{1}{1-a^2}}\int_0^{\pi/2}{\frac{dt}{\sqrt{(1-c^2
\sin^2t) [a^2/(1-a^2)+c^2\sin^2t)}}}, \label{final}
\end{equation}
where $c^2 =(b^2-a^2)/(1-a^2)$.
The integral $I_2$ is now of the form of $I_1$ and  (\ref{I2}) is obtained
after applying (\ref{I1}).


\end{document}